\documentclass[twocolumn]{aastex62}

\usepackage{amsmath}
\usepackage{color}
\usepackage{multirow}
\usepackage{graphicx}
\usepackage[toc,page]{appendix}
\usepackage{chemformula}

\begin{document}

\title{Two Terrestrial Planet Families With Different Origins}



\author{Mark R. Swain}
\affil{Jet Propulsion Laboratory, California Institute of Technology, 4800 Oak Grove Drive, Pasadena, California 91109, USA}
 
\author{Raissa Estrela}
\affil{Jet Propulsion Laboratory, California Institute of Technology, 4800 Oak Grove Drive, Pasadena, California 91109, USA}
\affil{Center for Radio Astronomy and Astrophysics Mackenzie (CRAAM), Mackenzie Presbyterian University, Rua da Consolacao, 896, Sao Paulo, Brazil}
 
\author{Christophe Sotin}
\affil{Jet Propulsion Laboratory, California Institute of Technology, 4800 Oak Grove Drive, Pasadena, California 91109, USA}

\author{Gael M. Roudier}
\affil{Jet Propulsion Laboratory, California Institute of Technology, 4800 Oak Grove Drive, Pasadena, California 91109, USA}
 
\author{Robert T. Zellem}
\affil{Jet Propulsion Laboratory, California Institute of Technology, 4800 Oak Grove Drive, Pasadena, California 91109, USA}


\begin{abstract}

The potentially important role of stellar irradiation in envelope removal for planets with diameters of $\lessapprox$ 2 R$_{\Earth}$ has been inferred both through theoretical work and the observed bimodal distribution of small planet occurrence as a function of radius.  We examined the trends for small planets in the three-dimensional radius-insolation-density space and find that the terrestrial planets divide into two distinct families based on insolation. The lower insolation family merges with terrestrial planets and small bodies in the solar system and is thus Earth-like. The higher insolation terrestrial planet family forms a bulk-density continuum with the sub-Neptunes, and is thus likely to be composed of remnant cores produced  by photoevaporation. Based on the density-radius relationships, we suggest that both terrestrial families show evidence of density enhancement through collisions. Our findings highlight the important role that both photoevaporation and collisions have in determining the density of small planets.

\end{abstract}


\section{Introduction} 

One of the important questions in the study of exoplanets  \citep{sotin2013} is ``Are terrestrial exoplanets Earth-like, Venus-like, or the remnants of gas- or ice-giants?''  Given sufficient atmospheric heating, a condition that is  met for many close-orbiting planets, theoretical studies \cite{owen2013,lopez13,lopez2014,rogers2015} predicted  photoevaporation of sub-Neptune H$/$He envelopes could create rocky super-Earth planets.  Potentially coexisting with photoevaporation, envelope loss due to core cooling has been proposed \citep{gupta2018}. With the original sub-Neptune envelope largely or completely removed through photoevaporation or another process, the planet radius in this scenario is determined by the size of a rocky, or possibly icy, core.  Recent work by \cite{fulton2017} demonstrated the presence of a deficit, or ``gap'',  in the planet occurrence rate and provided a compelling observational motivation for invoking the {\bf loss} of sub-Neptune H/He envelopes, potentially due to photoevaporation, as a formation mechanism for super-Earths. There are also indications that close orbiting transiting planets are larger around young stars, further indicating possible photoevaporation \citep{david2018}. To improve our understanding of the processes that shape the bulk properties of small planets, we explore the relationships between radius, insolation, and density for exoplanets and small bodies in our solar system.

\section{Methods and Results}

For this study, we used data from the NASA Exoplanet Archive and restricted the sample to confirmed planets with radius values $\leq$3.5 R$_{\Earth}$, augmented  with planets from \citet{marcy2014} and recent results for the Trappist-1 system \citep{grimm2018}. Kepler target radii were updated with GAIA-derived values  \citep{berger2018}. We removed planets with anomalously large densities ($\rho>15$ g/cm$^{3}$), a planet pair thought to have survived stellar engulfment \cite{charpinet2011},  and a planet pair with an anomalously large inclination difference, potentially indicating an unusual dynamical history \citep{rodriguez2018}, resulting in a sample of 107 exoplanets. In cases where planets had reported densities without uncertainties we used twice the average density uncertainty of the sample. To divide the sub-Neptunes from the terrestrials, we used a value of 1.75 R$_{\Earth}$ \citep{lopez2014}, which is also the approximate center of the occurance rate deficit identified by \cite{fulton2017}. Following \cite{zeng2017} and \cite{fulton2018}, we used insolation as a proxy for the photoevaporation capability of the host star, and we estimated the insolation using the method of \cite{weiss2014}. In cases where the semi-major axis values were not reported, the values were estimated from the period, assuming a circular orbit. The range of insolation for planets in this sample is greater than three orders of magnitude. When insolation is considered, we find that terrestrial planets separate into two families in the radius-insolation plane (see Figure 1), one with relatively lower levels of insolation (S/S$_{\Earth} < 10$) and one with higher levels of insolation (S/S$_{\Earth} > 10$), identified here  respectively as the T1 and T2 families.

\begin{figure}[htp!]
\centering
\includegraphics[width=1\columnwidth]{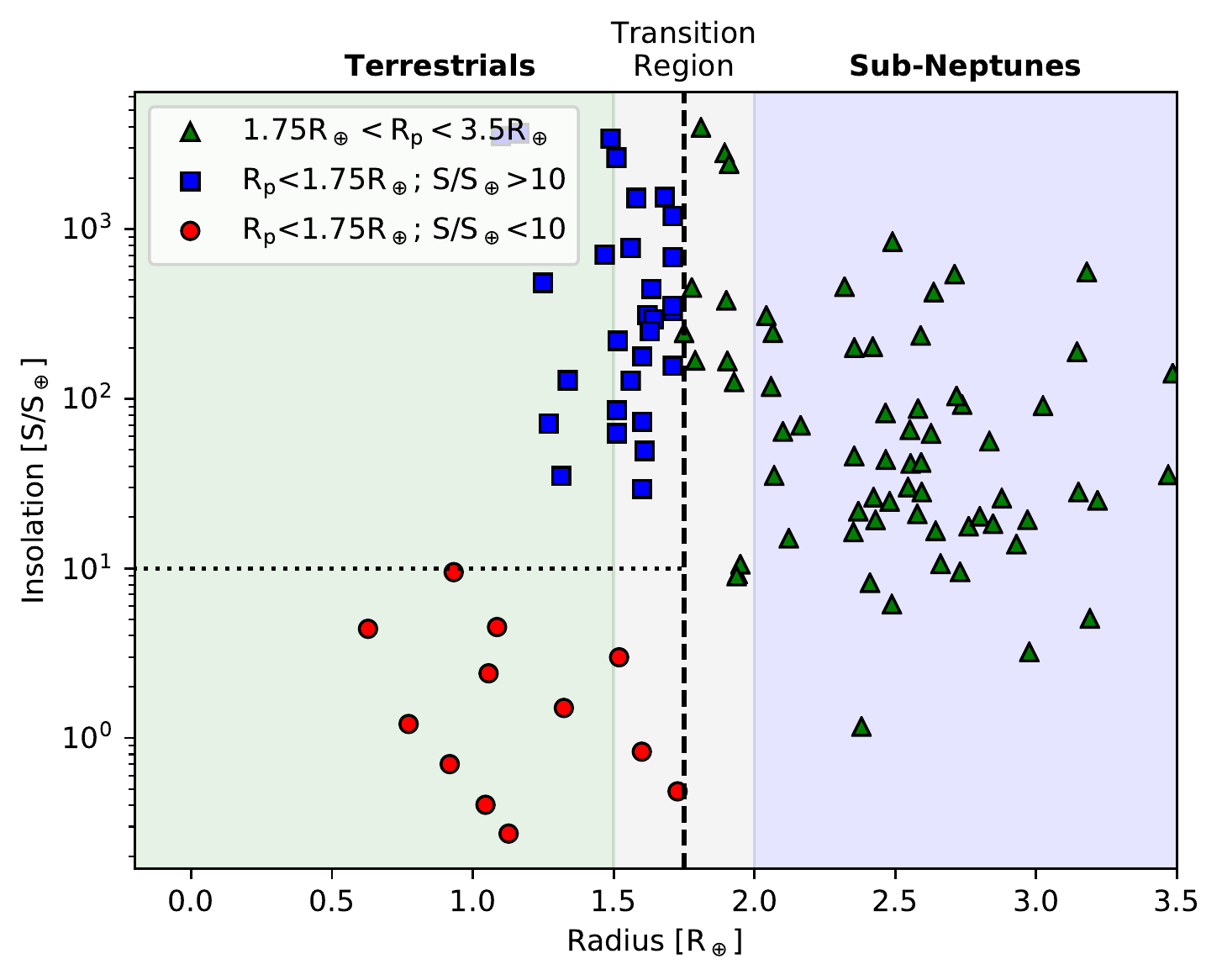}
\caption{An insolation gap divides the terrestrial planets into two categories. This division has a good correspondence to trends in the density-radius plane, and we use it to define two families for terrestrial planets. The low and high insolation terrestrial families (red circles and blue squares, respectively) are shown together with the sub-Neptunes (green triangles).}
\label{fig:CO}
\end{figure}

Interestingly, the separation of the T1 and T2 terrestrial planet families remains largely true in the radius-density plane. Additionally, photoevaporation of sub-Neptune (SN) H$/$He envelopes implies there should also be a connection between insolation and density. To explore that possibility, we constructed a radius-density diagram (see Figure 2 and also Figures 4 and 5). SN planets with radii $>1.75$ R$_{\Earth}$ have lower levels of insolation  than the T2 planets and are arranged as a consistent trend in the radius-density plane. For planets with radii 1.2 R$_{\Earth} <$R$<1.75$ R$_{\Earth}$, the average level of insolation increases. 

If we temporarily exclude the T1 terrestrial family from consideration, a general trend emerges of increasing planet density with decreasing planet radius. We modeled this trend with  a bilinear, piece-wise continuous function and retrieved the model posterior distributions using a Markov Chain Monte Carlo \citep{salvatier2016} method (see Figures 2 and 3). The average slope of the radius density function is larger for the T2 terrestrials (m$_{\rm{T2}}$=-12.17$^{+1.04}_{-1.03}$) than for the sub-Neptunes (m$_{\rm{SN}}$ = -2.31$\pm 0.08$. We determined that modeling the T1 and T2 planets separately was justified. The T1 sample slope is based on a fit where the average exoplanet density uncertainty is assigned to the solar system objects. For the T2 planet sample, both a $\chi^2$ test and the Akaike Information Criteria show a strong preference ($> 97.5$ \& $99 \%$, respectively) for rejecting the T1 model in favor of the T2 model. To test our selection of R$=1.75$ R$_{\Earth}$ for dividing the terrestrials from sub-Neptunes, we repeated the MCMC retrieval of parameters for a bilinear, piece-wise continuous model, but with the added degree of freedom that the bilinear intersection point was allowed to vary. We found the most probable value for the intersection point, which defines the boundary between terrestrials and sub-Neptunes in our model, was 2.28$^{+0.19}_{-0.07}$ R$_{\Earth}$. Additionally, \cite{martinez2019} identify a value of 2 R$_{\Earth}$ as a transition between terrestrials and sub-Neptunes. Taken together, this suggests that our T2 sample is not contaminated by the presence, of sub-Neptune type planets and, as we discuss later, there is additional evidence to support this conclusion as well.

The T1 terrestrial family follows a completely different density-radius trend with density increasing as a function of radius.  To probe the radius-density relationship between exoplanets and objects in our solar system, we assembled a list of 28 solar system bodies with reported densities and radii between  400 km $<$ R $<$ R$_{\Earth}$ (tabulated in the Appendix) that include terrestrial planets, moons, asteroids, and trans-Neptunian objects. When these solar system bodies are included in the radius-density plane, they form a continuous population with the  T1 family of low insolation terrestrial exoplanets, and we fit a linear model (m$_{\rm{T1}}$ = 3.01$\pm 0.32$) to the radius-density function of this family (see Figure 2).  This result for the density slope of the T1 exoplanets  plus solar system bodies is close to the prediction by \cite{sotin2007} for terrestrial planets (see Figures 2 and 4) as having a density relation of $\rho = 5.51$ R/R$_{\Earth}^{0.65}$ and seems to confirm the classification of the T1 family as telluric planets.

Our results are qualitatively consistent with the density-radius trend reported by \cite{weiss2014} but provide more detail and expand on the previous work in two important ways.  First, we incorporate the role of insolation and are thus able to identify specific subpopulations in the radius-density plane. Second, we are able to work with a significantly larger sample of both exoplanets and solar system objects. As we discuss below, this allows identification of two terrestrial planet families and a clearer picture of the sometimes competing, and sometimes complimentary, roles of collisions and photoevaporation in determining the properties of small planets.

\begin{figure*}[htp!]
\centering
\includegraphics[width=1\textwidth]{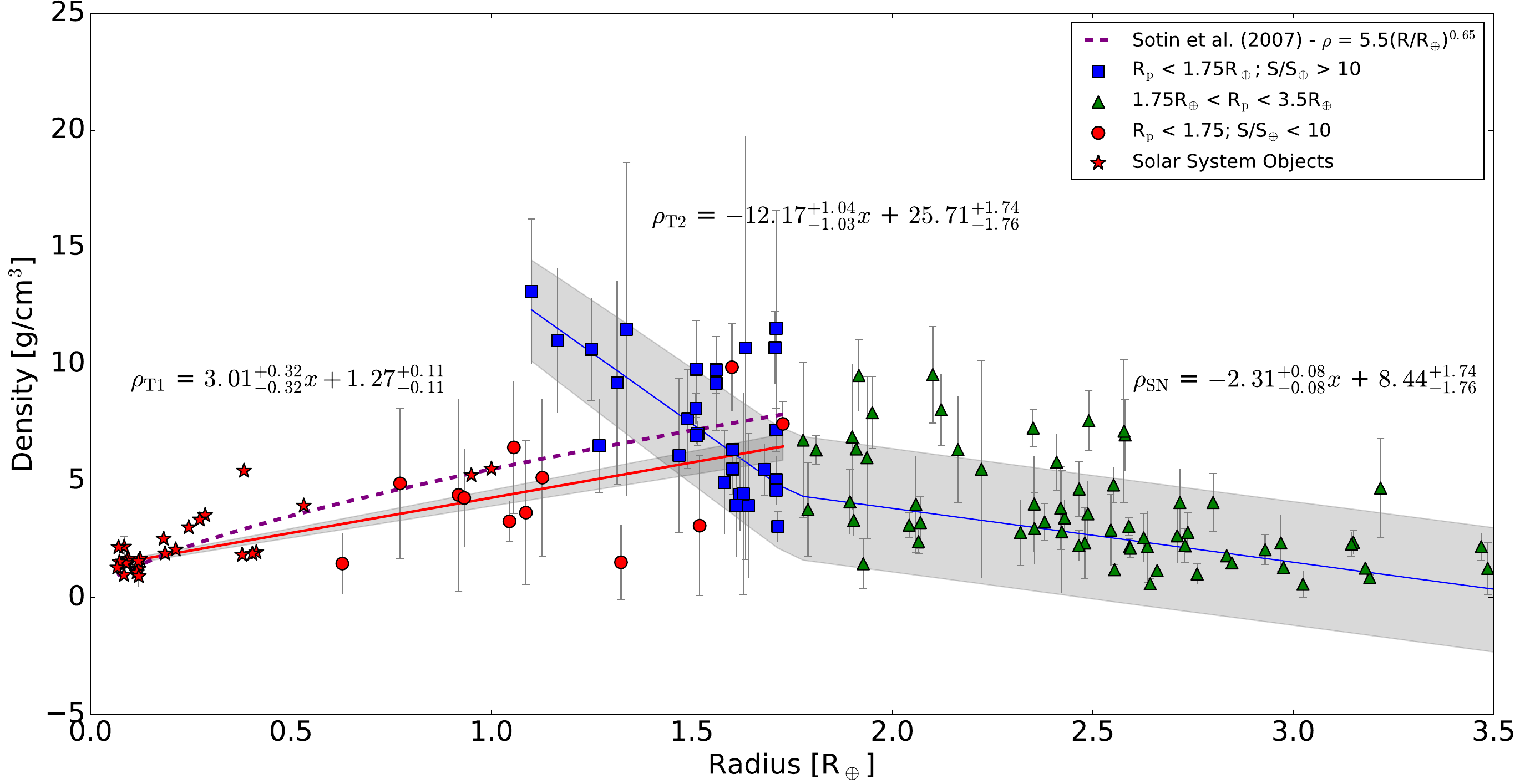}
\caption{The results of a bilinear fit to the high-insolation T2 terrestrials (blue triangles) and sub-Neptune exoplanets (green squares). A negative slope is detected at the $>5\sigma$ for both linear model components, implying photoevaporation is impacting planet density for both T2 terrestrial and sub-Neptune exoplanets. The low-insolation T1 terrestrial planets (red triangles) and solar system objects (stars) are fit with a linear model, which is in excellent agreement with previous theoretical modeling (dashed line) - see main body for further discussion. Data points are shown with $\pm 1 \sigma$ uncertainties, and the grey areas denote the $\pm 1 \sigma$ uncertainty regions for the linear models.} 
\end{figure*}

\begin{figure}[htp!]
\centering
\includegraphics[width=1\columnwidth]{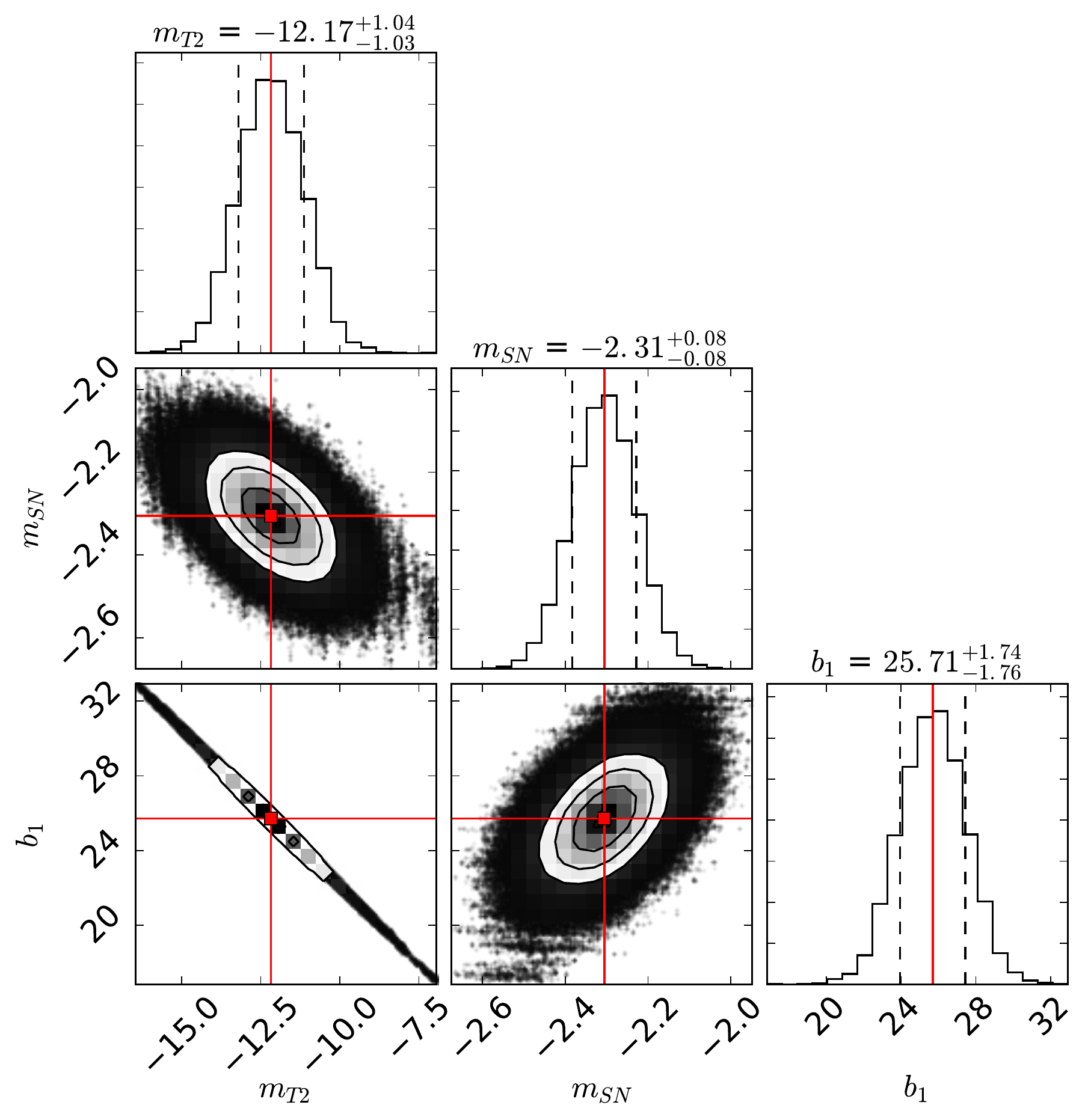}
\caption{Correlation plots and marginalized posterior distributions for the model parameters. The mean values are indicated in red and the $\pm 1 \sigma$ uncertainties are represented by the dashed lines.}
\label{fig:CO}
\end{figure}
%
\begin{table*}[htp!]
\centering
\caption{Density-Radius Slope Results}
\label{tab:atm models}
\begin{tabular}{|l|c|c|c|}
\hline
\textbf{Family/Region} &  \textbf{Domain} & \textbf{Density-Radius Slope} & \textbf{Average S/S$_{\Earth}$}\\
\hline
T1 terrestrials  & R $<$ 1.75R$_{\Earth}$, S/S$_{\Earth}<10$  &  m$_{\rm{T1}}$ = 3.01$\pm$0.32 & 2.43\\ 
\hline
T2 terrestrials &  R $<$ 1.75 R$_{\Earth}$, S/S$_{\Earth}>10$ & m$_{\rm{T2}}$=-12.17$^{+1.04}_{-1.03}$  & 804.4 \\ 
\hline
sub-Neptunes & 1.75 R$_{\Earth} <$ R $<$ 3.5 R$_{\Earth}$ & m$_{\rm{SN}}$ = -2.31 $\pm$0.08 & 256.1\\
\hline
transition region &  1.5 R$_{\Earth}$R$<$2.0 R$_{\Earth}$ & m$_{\rm{TR}}$ = -3.0$\pm$0.8 & 612.9\\ 
\hline
\end{tabular}
\end{table*}

\section{Discussion}
The density-radius relation for small planets and small bodies is an important diagnostic of the physical processes participating in planetary formation and evolution. While the likely role of photoevaporation in converting sub-Neptunes into super-Earth terrestrial planets is obvious in Figure 4, the slope of the density-radius function in the different domains is revealing. To assist in the interpretation of these findings, we will discuss each of the planet family groups identified in the previous section (T1s, T2s, and SNs) as well as a ``transition region'' identified as 1.5 R$_{\Earth}$ $<$2.0 R$_{\Earth}$; the density-radius slope and average insolation values for the planet families and the transition region are tabulated in Table 1.

\begin{figure*}
\centering
\includegraphics[width=1\textwidth]{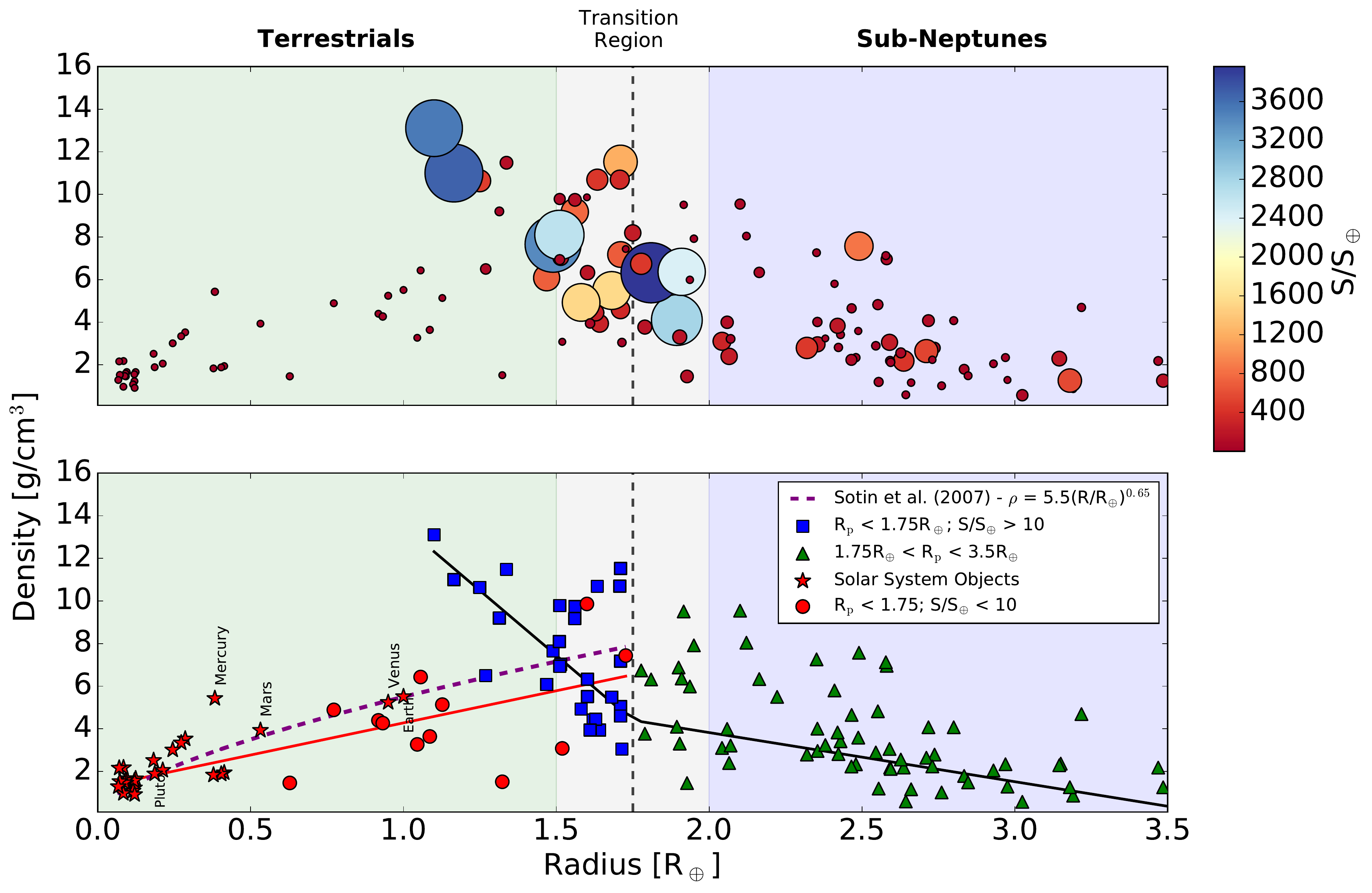}
\caption{Distinct trends in the density-radius and density-insolation planes separate the T1 (low insolation - red circles) terrestrials and the T2 (high insolation - blue squares) terrestrials. Collisions likely enhance the density of both the T1 and T2 families, while photoevaporation is likely modifying the density of sub-Neptunes (green triangles) and creating the T2 family by complete envelope stripping of some sub-Neptunes.}
\end{figure*}

{\bf Type 1 Terrestrials:} This family is characterized by comparatively low levels of insolation and a modest dependence of density on radius, with an average slope m$_{\rm{T1}}= 3.01\pm0.32$. A crucial aspect of this family is that it includes solar system terrestrial planets, exoplanet terrestrials, and solar system bodies that form  a continuous trend in an insolation-selected, radius-density plane. Within the radius-density plane, terrestrial planets in our own solar system are intermingled with the exoplanets, suggesting a common formation mechanism. Terrestrial planets in the solar system are believed to have formed through impacts (\cite{kokubo1998,schlichting2015,glenda2017} and references therein) and the continuity in the T1 family between exoplanets and solar system bodies with radii 400 km $<$R$<$ R$_{\Earth}$ is a strong indicator that assembly by collisions is important. The apparent continuity of the trend raises the question of whether the T1 family is consistent with oligarchic growth scenarios \citep{kokubo1998}. The modest, positive slope for the T1 density-radius function probably reflects a combination of density enhancement through compression, as the size of the bodies grow \citep{seager2007} and, potentially, loss of volatiles through collisions \citep{schlichting2015}.

{\bf Sub-Neptunes:} The pattern of increasing density with decreasing radius for the sub-Neptune family is consistent with modest levels of photoevaporation  from 2.0 R$_{\Earth} \leq$ R$\leq$ 3.5 R$_{\Earth}$ and the somewhat enhanced average insolation level in this region. If modest photoevaporation, in a statistical sense, plays a role in determining sub-Neptune density, it is removing of $\sim$ half the envelope mass ($\sim$1.5-3\% of the total planet mass), corresponding to a typical sub-Neptune radius reduction of $\sim35$\% for R$\sim$2.0 R$_{\Earth}$ or $\sim$ a quarter of the envelope mass, corresponding to a radius reduction of $\sim20$\% for R$\sim$3.0 R$_{\Earth}$ \citep{lopez2014}. 

{\bf Type 2 Terrestrials:} This family of terrestrial planets is characterized by high levels of insolation and a strong inverse dependence of density on radius with an average slope of the density-radius function of {m$_{\rm{T2}}= -12.17^{+1.04}_{-1.03}$}. At the smallest radii for this family, density values are high, $\sim$ 10 g$/$cm$^{3}$. These are extremely large values for terrestrial planet density and some explanation is required. One possibility is that the decompression triggered by the evaporation of the atmosphere is not instantaneous \cite{mocquet2014}. However, the negative slope of the density-radius function implies that gravitational compression cannot alone be responsible for the high density values. Theoretical studies indicate that T2 planets likely have had the envelope completely stripped and are``bare cores'' \citep{owen2017,jin2018,vaneylen2018}. While the bare core scenario may not apply in the case of a secondary atmosphere established by mantle outgassing \citep{dorn2018} or delivered in the form of volatiles associated with accreted planetesimals \citep{elkins-Tanton2008}, the slope of the T2 density-radius function is likely too steep to be consistent with these scenarios. Density enhancement of super-Earths by collisions has been studied \citep{marcus2009}, and we suggest a modification of this scenario in which envelope-stripped bare cores have undergone a significant level of planetesimal bombardment early in the planet's history; the planetesimal impacts raise volatile plumes that are then efficiently stripped by the high-insolation levels. This scenario is consistent with the correlation between enhanced density and insolation that is present in this sample (see Figure 5). However, photoevaporation is not responsible for density enhancement in the majority of T2 planets because of a positive correlation between density and mass (see Figure 6) that is consistent with density enhancement by collisions; systematic density enhancement due to photoevaporation would produce a slope of the opposite sign. The plausibility of density enhancement through collisions is also supported by recent observations \citep{bonomo2019}  consistent with theoretical predictions \citep{marcus2010b}. Other studies show that some super-Earths are expected to go through a giant impact phase after dispersal of the gas disk \citep{cossou2014,izidoro2017}.

Impact driven atmospheric removal has been studied in the context of Earth \citep{schlichting2015} and is likely more efficient for the significantly higher levels of insolation associated with T2 planets. In our proposed secnario, the combination of planetesimal impacts on a bare core, in the presence of strong insolation, produces a fractional distillation type of effect where the heavy elements are preferentially retained, or reaccreted, and lighter materials are vaporized and then stripped. This process is more efficient when the gravitational binding energy is lower, which is consistent with the negative slope for the T2 density-radius relation. Further, the reaccretion of impact-produced siderophile elements has the potential to act as a reducing agent and transform atmospheric CO$_{2}$-H$_{2}$O into H$_{2}$ \citep{glenda2017}, further accelerating H loss and the density enhancement process. Work by \cite{schlichting2015} shows this atmospheric loss process can also operate in the presence of an Earth-like atmosphere, implying that a bare core is not a requirement for a combined impact/photoevaporation-driven density enhancement process.

{\bf Transition Region:} Following \cite{lopez2014}, this region corresponds to the range of radii (1.5 R$_{\Earth} \leq$ R $\leq 2.0$ R$_{\Earth}$) associated with the transition from sub-Neptunes to super-Earths. In our sample, this region contains planets with densities that are indicative of sub-Neptunes and super-Earths, and it also corresponds to the planet occurrence deficit identified by \cite{fulton2017}. The transition region spans the sub-Neptune-terrestrial boundary and also contains the highest average level of insolation of any part of the radius-density plane.  Planets in this region, especially the lower density ones, offer the potential to observationally probe the process of envelope loss. The combination of T2 planets both inside and outside the transition region suggests there may be a wide diversity of timescales for envelope loss.  

The T1, T2, and SN planet families, and the transition region, highlight the importance of stellar radiation, in either its presence or absence, on the formation and evolution of small planets. Fundamentally, there are three important axes (radius, insolation, and density) that allow identification of separate populations, and these have the potential to become confused when projected into 2-dimensional spaces. We illustrate this in Figure 2. Considering the identification of the separate populations in a 3-dimensional space allows informed speculation about whether the T1 and T2 populations merge. Our prediction is that, as more terrestrial-type planets are found, the T1 and T2 populations will {\it appear} to merge in a density-radius plot, but, in actuality, they will remain separate and fundamentally distinct populations in a radius-insolation-density space. 

The initial identification of the T1 and T2 groups was based on insolation differences of planets with R$<$1.75 R$_{\Earth}$, but this single-parameter observational distinction reflects a much more profound difference. While the T1 and T2 planet families can be termed terrestrial because of the combination of their radii and their bulk densities, which imply they must be rocky, their density-radius relation suggests that they have fundamentally different formation histories. Our contention is that T1 objects are assembled``from the bottom up'' by a collisional process that assembles larger bodies from smaller pieces and hence, the T1 planets are truly Earth-like. In contrast, the T2 objects are ultimately produced from a ``top down'' evaporation of sub-Neptunes. Thus the T2 planets are the rocky core remnants of small gas giants that are linked, implied by the high insolation level of the T2 sample, to the photoevaporation of sub-Neptunes. 

The above interpretation of the T2 formation history implies that the T2 sample provides an opportunity to study the formation process of gas giant planets and directly probe the cores believed to be  responsible for the onset of gas accretion \cite{pollack96}.  However, care must be taken because the same cores that had their envelopes removed by photoevaporation could also have been further processed by additional stripping of the outer layers of the mantle. If mantle stripping were common in the T2 sample, it would be manifested as a inverse relationship between density and mass because the denser inner core would represent a larger fraction of the exoplanet. We tested this hypothesis (see Figure 6) and find exactly the opposite; we find a positive correlation between T2 planet mass and density. The implication of the T2 density-mass relation is photoevaporation is not signicantly processing the bare cores in the T2 sample. By extension, the T2 sample is not contaminated by SN planets. This finding is important; it means that the negative slope for the T2 planet density-radius relation cannot be caused by photoevaporation of mantle material. Thus, either some aspect of the initial core formation process for gas giants produces $both$ a positive correlation of mass and density $and$ a negative correlation of density and radius, or, collisional processes have a role in enhancing the mass and heavy element content of T2 planets.

In both the T1 and T2 terrestrials, collisions likely play an important role in density enhancement through driving off volatiles.  But photoevaporation has a completely unique role in the sculpting of the T2-SN density-radius relationship. If collisions do play a significant role in the density enhancement of some T2 objects, it implies their entire envelope formation and loss sequence happened extremely quickly, potentially allowing the super-Earth bare cores to undergo significant bombardment. Collisional processing, inferred from the positive density-mass relation for T2 planets,  implies that the average mass for the T2 sample, 4.8$\pm$1.8 M$_{\Earth}$,  likely represents an upper bound for the value typically needed to produce the onset of the rapid gas accretion phase associated with envelope assembly during the planet formation process. Our results are in agreement with core mass predictions by \cite{lee2016} and, in view of core mass estimates by \cite{batygin2016}, raise the question of whether sub-Neptune and hot-Jupiter cores form in the same disk environment. 

\begin{figure}[htp!]
\centering
\includegraphics[width=1\columnwidth]{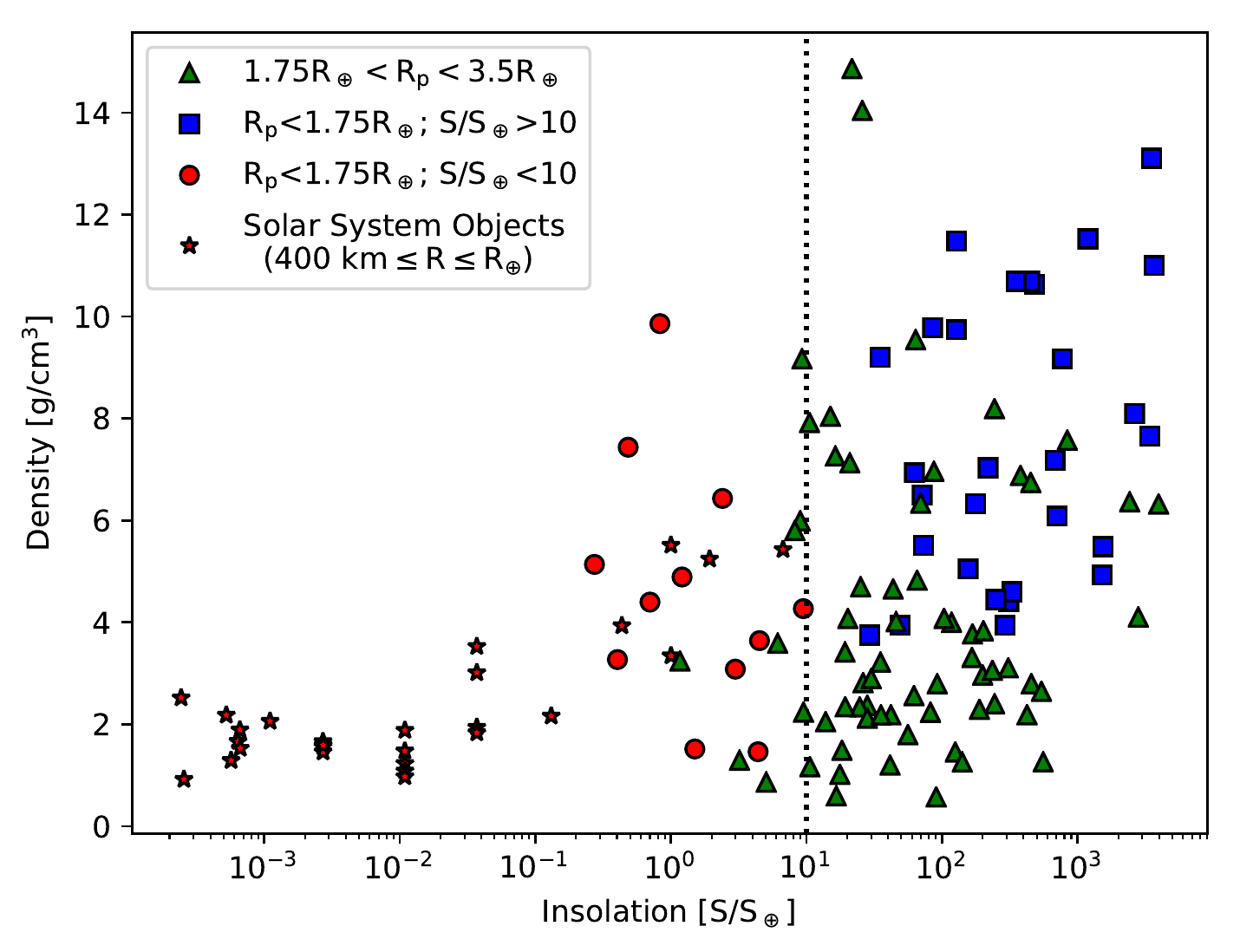}
\caption{Enhanced levels of insolation are correlated with enhanced planet densities. The plot symbols are defined as in Figure 4.}
\end{figure}

\begin{figure}[htp!]
\centering
\includegraphics[width=1\columnwidth]{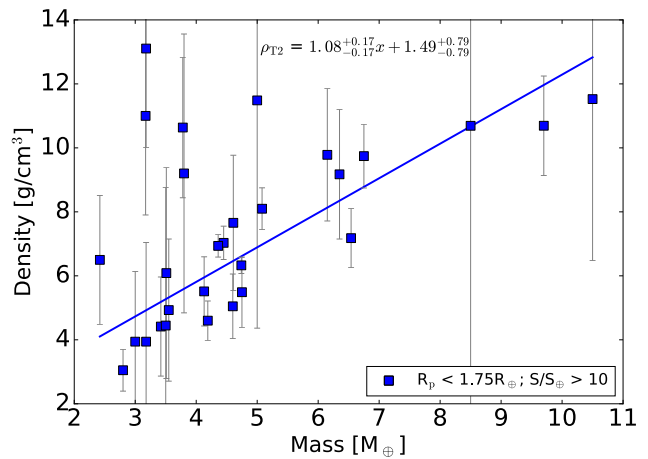}
\caption{The positive slope for the T2 terrestrial planets density-mass relationship is consistent with mass-density enhancement by collisions and is incompatible with density enhancement due to photoevaporation of mantle material which would yield a slope of the opposite sign. If collisions are indeed modifying the density and mass of these planets then the T2 average mass and density values may represent upper limits for the core mass needed to initiate envelope accretion.}
\end{figure}

\section{Conclusions}

When considering a combination of exoplanets with known densities and radii 0$<$R$<$ 3.5 R$_{\Earth}$, together with solar system bodies with known densities and radii 400 km $<$R$<$ R$_{\Earth}$, and when considering insolation, we find the following. Terrestrial planets naturally divide themselves into two families that each form a continuous trend in the radius-density space. The T1 terrestrial family includes both exoplanets and the terrestrial planets and small bodies in our solar system. The density-radius-insolation relation for the T1 family is consistent with assembly of these bodies through collisions. Relatively low levels of insolation ($<$ 10 S/S$_{\Earth}$) are characteristic of the T1 family.

A second terrestrial family, the T2 family is marked by relatively high levels of insolation, includes only exoplanets, and has a density-radius trend that is piece-wise continuous with the sub-Neptunes. The T2 density-radius trend implies that collisions, in the presence of strong photoevaporation, can create extremely high-density terrestrial planets ($\rho\sim10$ g$/$cm$^{3}$).  A high insolation transition region overlaps the junction between the T2 super-Earths and the sub-Neptunes. The sub-Neptune density-radius relation also shows evidence of modest photoevaporation. Taken together, the sub-Neptunes, transition region, and T2 super-Earths indicate the pervasive role of photoevaporation in sculpting a continuum from low density ($\rho \leq 1$ g$/$cm$^{3}$) planets with large H/He envelopes to extremely high-density terrestrials ($\rho \sim 10 $g$/$cm$^{3}$). The strong implication is that T2 super-Earths are the remnant cores of small gas giant planets and were created by photoevaporative stripping of sub-Neptunes. However, the T2s are created not only by photoevaporation/stripping; the T2 mass-density correlation implies a process which increases mass with increasing density, potentially late bombardment of naked cores. The potential role of collisions in creating the high density T2 planets implies that the process of envelope assembly and stripping must be rapid for the highest density T2 planets, although the envelope stripping timescale could be much slower for T2 planets in the transition region. 

Finally, we are in a position to provide an observations-based answer to the question posed in the introduction, ``Are terrestrial exoplanets Earth-like, Venus-like, or the remnants of gas- or ice-giants?''   Invoking the simplification that Earth-like and Venus-like are essentially the same, we can answer ``both''. Terrestrial planets apparently form from two mechanisms. One mechanism is ``Earth-like'' and relies on terrestrial planet formation by collisions. The other mechanism is through photoevaporation, which produces remnants of gas- or ice-giants.

\section*{Acknowledgments}

We thank Caitlin Griffith, Yasuhiro Hasegawa, and Konstantin Batygin for helpful discussions and encouragement in the preparation of this manuscript. This research has made use of the NASA Exoplanet Archive, which is operated by the California Institute of Technology, under contract with the National Aeronautics and Space Administration under the Exoplanet Exploration Program. This research has also made use of the JPL Solar System Dynamics database. Raissa Estrela acknowledges Sao Paulo Research Foundation (FAPESP) for the fellowship \#2018/09984-7. This work has been supported in part by the California Institute of Technology Jet Propulsion Laboratory Exoplanet Science Initiative. This research was carried out at the Jet Propulsion Laboratory, California Institute of Technology, under a contract with the National Aeronautics and Space Administration.

\bibliography{references}


\appendix

Table of solar system bodies, with radius R$_{\Earth}$ to 400 km. This list of 28 solar system bodies with reported densities includes the terrestrial planets, moons, minor planets, asteroids, and trans-Neptunian objects.  The majority of the density data is taken from the JPL Solar System Dynamics (SSD) database augmented by references in the literature when necessary. For objects that are planetary satellites, we use the planet semi-major axis for calculating insolation.


\begin{table}[htp!]
\centering
\begin{tabular}{l|ccc|r}
\tablecaption{Solar System Bodies}
\textbf{Body} & \textbf{Mean Radius (km)} & \textbf{Mean Density (g/cm3)} & \textbf{Semi-Major Axis (au) } & \textbf{Citations} \\ 
\hline
Earth 			&  6378.1366$\pm$0.0001		& 5.5136$\pm$0.0003 		&  $a_{E}$ 	& SSD\\
Venus		 	& 6051.8$\pm$1.0	& 5.243$\pm$0.003	& 	$a_{V}$	& SSD \\
Mars 			& 3396.19$\pm$0.1	&	3.9341$\pm$0.0007		&	$a_{M}$	& SSD \\
Ganymede 	& 2631.2$\pm$1.7	&	1.942$\pm$0.005	&	$a_{J}$	& SSD \\
Titan 			& 2574.73$\pm$0.09	&	1.882$\pm$0.001 &	$a_{S}$	& SSD \\
Mercury 		& 2440.53$\pm$0.04	&	5.4291$\pm$0.0007	&	$a_{Mer}$	& SSD \\
Callisto 		& 2410.3$\pm$1.5 	&	1.834$\pm$0.004 &	$a_{J}$	 & SSD \\
Io 					& 1821.6$\pm$0.5	&	3.528$\pm$0.006 &	$a_{J}$ & SSD \\
Moon 			& 1737.5$\pm$0.1 	&	3.344$\pm$0.005	&	$a_{E}$	& SSD \\
Europa 			& 1560.8$\pm$0.5	&	3.013$\pm$0.005 &	$a_{J}$ & SSD \\
Triton 			& 1353.4$\pm$0.9	&	2.059$\pm$0.005	&	$a_{N}$	& SSD \\
Pluto 			& 1188.3$\pm$1.6		&	1.89$\pm$0.06	&	$a_{P}$	& SSD \\
Eris 				& 1163$\pm$6	&	2.52$\pm$0.05 & 67.74049521464768$\pm$0.0027096 & [1], SSD \\
Humea 			& 797.5$\pm$5 &	1.89$\pm$0.08	&	43.3 & [2], [3] \\
Titania 			& 788.9$\pm$1.8	&	1.662$\pm$0.038	 &	$a_{U}$ & SSD \\
Rhea 			& 764.30$\pm$1.10	&	1.233$\pm$0.005		&	$a_{S}$ & SSD \\
Oberon			& 761.4$\pm$2.6	&	1.559$\pm$0.059 	&	$a_{U}$	& SSD \\
Iapetus 		& 735.60$\pm$1.50	&	1.083$\pm$0.007	&	$a_{S}$	& SSD \\
2007 OR10 	& 767.5	$\pm$112.5 &	0.92$\pm$0.46	&	67.37610770137752$\pm$0.0073504	& [4], SSD \\
Charon 			& 603.6$\pm$1.4	&	1.664$\pm$0.012	&	$a_{P}$	&  SSD \\
Umbriel 		& 584.7$\pm$2.8	&	1.459$\pm$0.092	&	$a_{U}$	& SSD \\
Ariel 				& 578.9$\pm$0.6	&	1.592$\pm$0.092	&	$a_{U}$	&  SSD \\
Dione 			& 561.70$\pm$0.45	&	1.476$\pm$0.004	&	$a_{S}$	& SSD \\
Quaoar 			& 535$\pm$19	&	2.18$\pm$0.43	&	43.69157469300723$\pm$0.0022972	& [5], SSD \\
Tethys 			& 533.00$\pm$0.70	&	0.973$\pm$0.004 &	$a_{S}$ & SSD \\
Ceres 			& 445.6$\pm$1.0	& 2.162±0.0008	&	2.769165146349478$\pm$2.5823e-11	& [6], SSD \\
Orcus 			& 459$\pm$13	& 1.53$\pm$0.15	&	39.26631237879172$\pm$0.00090814 & [5], SSD \\
Salacia 			& 427$\pm$23	& 1.29$\pm$0.29	&	42.05690689579897$\pm$0.0056705 & [5], SSD \\
\end{tabular}
\label{tab:system_properties}
\caption{Semi-major axis values for the less commonly known solar system bodies are listed. For determining the insolation values of the moons, we use the orbit semi-major axis value for the parent body. The SSD site can be accessed at https://ssd.jpl.nasa.gov/. Reference key: [1] \cite{sicardy2011}, [2] \cite{ortiz2017}, [3]  \cite{rabinowitz2006}, [4] \cite{kiss2018}, [5] \cite{fornasier2013}, [6]  https://web.archive.org/web/20150905125337/http://nesf2015.arc.nasa.gov/sites/default/files/downloads/pdf/05.pdf.}
\end{table}

\end{document}